\documentstyle[11pt,twoside,epsfig]{article}

\newlength{\dinwidth}
\newlength{\dinmargin}
\newcommand{\begl}{\begin{equation}}
\newcommand{\engl}{\end{equation}}

\def\TeV{\hbox{$\;\hbox{\rm TeV}$}}
\def\GeV{\hbox{$\;\hbox{\rm GeV}$}}

\def\m{\hbox{$\;\hbox{\rm m}$}}
\def\cm{\hbox{$\;\hbox{\rm cm}$}}

\def\pb{\hbox{$\;\hbox{\rm pb}$}}

\begin{document}
\begin{titlepage}
%
%
\noindent
{\tt DESY 96-163    \hfill    ISSN 0418-9833} \\
\vspace*{4.cm}
\begin{center}
\begin{Large}
%
%
{\bf Search for Excited Fermions with the H1 Detector}\\
\vspace*{2.cm}
H1 Collaboration \\
\end{Large}
\vspace*{4.cm}
{\bf Abstract:}
\begin{quotation}
%
%
We present a search for excited electrons, neutrinos and quarks 
using the H1 detector at the $ep$ collider HERA, 
based on data taken in 1994 with an integrated luminosity
of 2.75 pb$^{-1}$. 
Radiative decays of excited quarks and neutrinos have been
investigated as well as decays of excited electrons into
all possible electroweak gauge bosons. No evidence for new particle
production is found and exclusion limits are derived.

\end{quotation}
\vfill
%
%
\cleardoublepage
\end{center}
\end{titlepage}
\begin{flushleft}
 S.~Aid$^{13}$,                   
 M.~Anderson$^{23}$,              
 V.~Andreev$^{26}$,               
 B.~Andrieu$^{29}$,               
 A.~Babaev$^{25}$,                
 J.~B\"ahr$^{36}$,                
 J.~B\'an$^{18}$,                 
 Y.~Ban$^{28}$,                   
 P.~Baranov$^{26}$,               
 E.~Barrelet$^{30}$
% ...

\end{flushleft}
\begin{flushleft} {\it
 $ ^1$ I. Physikalisches Institut der RWTH, Aachen, Germany$^ a$ \\
% ...
$ ^{13}$ II. Institut f\"ur Experimentalphysik, Universit\"at Hamburg,
                                     Hamburg, Germany$^ a$  \\
% ...
$ ^{18}$ Institute of Experimental Physics, Slovak Academy of
                Sciences, Ko\v{s}ice, Slovak Republic$^{f, j}$\\
% ...
$ ^{23}$ Physics Department, University of Manchester,
                                          Manchester, UK$^ b$\\
% ...
$ ^{25}$ Institute for Theoretical and Experimental Physics,
                                                 Moscow, Russia \\
 $ ^{26}$ Lebedev Physical Institute, Moscow, Russia$^ f$ \\
% ...
$ ^{28}$ LAL, Universit\'{e} de Paris-Sud, IN2P3-CNRS,
                            Orsay, France\\
 $ ^{29}$ LPNHE, Ecole Polytechnique, IN2P3-CNRS,
                             Palaiseau, France \\
 $ ^{30}$ LPNHE, Universit\'{e}s Paris VI and VII, IN2P3-CNRS,
                              Paris, France \\
% ...
$ ^{36}$ DESY, Institut f\"ur Hochenergiephysik,
                              Zeuthen, Germany$^ a$\\
% ...
$ ^a$ Supported by the Bundesministerium f\"ur Bildung, Wissenschaft,
        Forschung und Technologie, FRG,
        under contract numbers 6AC17P, 6AC47P, 6DO57I, 6HH17P, 6HH27I,
        6HD17I, 6HD27I, 6KI17P, 6MP17I, and 6WT87P \\
 $ ^b$ Supported by the UK Particle Physics and Astronomy Research
       Council, and formerly by the UK Science and Engineering Research
       Council \\
% ...
$ ^f$ Supported by the Deutsche Forschungsgemeinschaft \\
% ...
$ ^j$ Supported by VEGA SR grant no. 2/1325/96 \\

   } \end{flushleft}
%
\newpage
\section{Introduction}
In composite models of quarks and leptons it is natural to expect
new heavy particles, which can be interpreted as excitations of
the ground state.
The direct observation of either excited leptons or excited 
quarks would certainly provide strong evidence for a 
new layer in the substructure of matter.

Electron proton interactions at very high energies provide
an excellent environment to look for  
excited leptons and quarks. If we consider for example excited electrons,
these states can only be reached
via magnetic type couplings~\cite{Low}
because the electromagnetic current is conserved
in electromagnetic interactions.  
It has been 
shown \cite{Hagiwara},
that about half of the cross section is expected in the elastic channel
\begl
e+p\rightarrow e^*+p
\engl
with a  branching ratio of e.g. $30\%$~\cite{Zerwas2}
into $e\gamma$ for the decay of an $e^*$ with a 
mass of more than $150$ GeV. 
This reaction has an extremely clear signature,
a wide angle electron photon pair and nothing else in the detector.
The analysis of other decay channels (e.g. $e^*\rightarrow
e+jets$) or of inelastically produced excited electrons 
($e+p\rightarrow e^*+X$) is more complicated.
The present paper covers elastic and inelastic $e^*$ production in
all decay channels. 

The magnetic coupling can also be applied to other gauge boson interactions.
Excited neutrinos could for example be produced in charged current
reactions, $e+p\rightarrow \nu^*+X$. 
The $\nu^*$ will decay into $\nu\gamma,\nu Z$ or $\nu W$. This analysis
considers a search for 
$e+p\rightarrow \nu^*+X, \nu^*\rightarrow \nu+\gamma$  
because of the clear signature of this particular decay
channel and because
theoretical models~\cite{Zerwas2} can be constructed which predict 
a high branching ratio $BR^*$ of $\nu^*\rightarrow\nu\gamma$. 

In the same way the magnetic transition coupling of quarks 
would allow single production of excited quarks
through $t$-channel gauge boson exchange between the incoming
electrons and quarks. The cross section
is dominated by $\gamma$ exchange with
very small values of the squared momentum transfer (photoproduction limit) 
and therefore the scattered electron is unseen in the detector.
The decay $q^*\rightarrow q+\gamma$ again provides
a clear signature and a search for this channel is also described
in this paper.

The essential parameter in the production and decay of excited fermions ($f^*$)
of mass $M$ is 
the partial width $\Gamma(f^*\rightarrow fV)$, where $V$ represents
a heavy boson ($W,Z$) or photon. In the photonic case this partial
width is given by
\begl
\Gamma(f^*\rightarrow f\gamma)=\frac{\alpha M^3 c^2_{\gamma f^*f}}{\Lambda ^2}
\enspace .
\engl
A general treatment can be found in ref.~\cite{Zerwas2}.
The  couplings $c_{Vf^*f}$ are a priori unknown. The theoretical
calculation is model dependent, see for example~\cite{Zerwas2}. 
The parameter $\Lambda$ determines the compositeness
scale. It is presently constrained to values above 0.7 to 1 TeV. 

In the narrow width approximation the total cross section for  elastic
production of excited electrons  is given by
\begl
\sigma(ep\rightarrow e^*p)=\frac{4\pi^2}{s}(2J+1)\frac{\Gamma}{M}
f_{\gamma/p}(M^2/s)
\engl
where $J$ is the spin of the new fermion and $\sqrt{s}$ the center
of mass energy ($\approx 300$ GeV at HERA).
The number of photons with momentum fraction between $z$ and
$z+dz$ radiated off the
proton is given by $f_{\gamma/p}(z)$. 
It can be evaluated using e.g.~the formula given in ref.~\cite{Kniehl}.
In the case of excited quark
production the relevant formula includes a convolution of
the photon densities $f_{\gamma/e}$ in the electron and
the quark densities $f_{q/p}$ in the proton. 
More general expressions for $e^*$ and  $\nu^*$ production are given in
ref.~\cite{Hagiwara}.
Because the calculation of the widht, $\Gamma$, and the branching ratio,
$BR^*$, for the decay of the new fermion into a specific
final state are model dependent, the experimental exclusion limits obtained
in this paper
are quoted for the products $\sigma\cdot BR^*$. They can therefore also be used
as exclusion limits for arbitrary heavy objects decaying into the
investigated final states.

Although most models
would expect excited states with masses of the order of the compositeness
scale it is important to cover experimentally  the
$M<300$ GeV region accessible for direct searches at HERA. This analysis
uses 
data taken with the H1 detector in 1994 at HERA with
an integrated luminosity of 2.75 pb$^{-1}$ in positron proton
collisions\footnote{Only for $\nu^*$ production
the charge of the incoming lepton is relevant.}
at a center of mass energy of 300 GeV. Previous
searches for excited fermions made 
with the H1-detector are described in ref.s~\cite{bib.h1_93}
and~\cite{bib.h1_94}. A similar search has been presented by the
ZEUS collaboration~\cite{ZEUS}.

\section{Experimental set-up}

A detailed description of the H1 detector can be found
elsewhere~\cite{bib.h1detector}. Here we describe briefly the components
which are relevant for this analysis.
The interaction region is surrounded by a drift and
proportional chamber tracking system, split into central
barrel and forward parts and covering the angular range
7$^{\circ} \le \theta \le$ 176$^{\circ}$
\footnote{The forward direction (positive $z$-coordinate)
          from which the polar angle, $\theta$,
          is measured is the
          proton beam direction.}.
The tracking system is placed inside a
finely segmented
liquid argon (LAr) calorimeter covering the range 4$^{\circ} \le
\theta \le 153^{\circ}$. The electromagnetic section,
between 20 and 30 radiation lengths ($X_0$) deep, is used to
identify electrons and photons. Hadrons also deposit energy in
the outer layers (hadronic section) of the LAr calorimeter.
The total thickness of
the LAr components varies between 4.5 and 8 interaction lengths.
For the LAr calorimeter energy resolutions of
$\sigma(E)/E \simeq$ $12$ \%/$\sqrt{E\, (\mbox{GeV})} \oplus 1\%$
for electrons and
$\sigma(E)/E \simeq$ $50$ \%/$\sqrt{E\, (\mbox{GeV})} \oplus 2\%$
for hadrons have been obtained in test beam measurements.
A 22 $X_0$ deep lead-scintillator electromagnetic calorimeter (BEMC)
is located in the backward region (151$^{\circ} \le \theta \le$
176$^{\circ}$) of the H1 detector. The resolution is determined to be
$\sigma(E)/E \simeq$ $10$ \%/$\sqrt{E\, (\mbox{GeV})} \oplus 2\%$.\par

The tracking system and calorimeters are surrounded by a
super-conducting solenoid producing a uniform field of $1.15$
T in the $z$-direction.
The iron return yoke of the main solenoid
is instrumented with limited streamer tubes
to measure tracks of penetrating muons
in the range of polar angles
$5^{\circ} \le \theta \le 170^{\circ}$.
There is an additional forward muon spectrometer 
to measure high energy muons in the region 
$3^{\circ} \le \theta \le 17^{\circ}$.\par

The luminosity detectors, which measure
the Bethe-Heitler reaction $e p \rightarrow e \gamma p$ and
electrons from photoproduction processes, are placed at the
positions $z = -33\m$ (electron tagger) and at $z = -103\m$
(photon tagger), measured along the direction of
the proton beam with respect to the nominal interaction point.

\section{Search for excited leptons}

\subsection{Exclusive electromagnetic cluster analysis}
 
The first search method looks for $e\gamma$ pairs in an otherwise
empty detector. Exotic candidates for this reaction are elastically
produced $e^*$s decaying into an electron and a photon. Also
included are the so called quasi-elastic reactions
(e.g. $e+p\rightarrow e^*+\Delta^+$) where the decay products of
the hadronic state remain in the beam pipe. 

Events have been selected with only two electromagnetic clusters  
and little additional energy observed in the detector.
The main background source stems from the Wide-Angle
Bremsstrah\-lung (WAB) process $ep \rightarrow e \gamma p(X)$
with a well separated $e\,\gamma$ pair in the final state.
The WAB process constitutes a kinematically
indistinguishable background to $e^*$ production.  Optimizing the
cuts for the selection of WAB events therefore also optimizes the
efficiency for excited electrons decaying into $e+\gamma$.

Most of the WAB events are 
expected to have an $e\, \gamma$ system of low invariant
mass with high momentum in
the direction of the incoming electron. 
The system is boosted into the forward direction 
only for masses of the $e\, \gamma$ system above $2\,E$, where $E$ is
the energy of the incoming electron (27.5 GeV). 

Another type of expected background is that  due to the so called
Two Photon
($\gamma\, \gamma$) process
$ep \rightarrow e e e p(X)$,
where an electron positron pair is produced in addition to the
scattered positron. If two of these three leptons are seen inside
the detector and one of them has no reconstructed track, due to
geometrical acceptance or chamber inefficiencies, the event
could be misinterpreted as an $e\, \gamma$ candidate.

The cuts to select $e\, \gamma$ pairs are shown in
table~\ref{tab.ls.sel.wab}.
The events are required to have satisfied the BEMC or LAr
energy triggers. Trigger efficiencies have been studied
using selected WAB events with one electromagnetic cluster in the BEMC
and the other in the LAr calorimeter. The trigger efficiencies
are $\approx 100\,\%$ for electromagnetic clusters with energies above
$8\GeV$ in the BEMC, above $14\GeV$ in the LAr barrel region
($\theta > 20^{\circ}$), and above $30\GeV$ in the LAr forward
region ($\theta < 20^{\circ}$). Parameterized trigger efficiencies
were then folded into the Monte Carlo  models for the simulation of
$e^*$ production.

\begin{table}[htb]
\begin{center}

\begin{tabular}{l}
 \hline
 1. $ \geq 2$ isolated e.m.\ clusters, $E_i > 2\GeV$
       for $i=e, \gamma$ \\
 
 2. $ E_e + E_{\gamma} > 20\GeV$  \\
 
 3. $ E_{total} - E_e - E_{\gamma} < 5\GeV  $   \\
 
 4. $ |m(\theta_e,\theta_{\gamma})-m| < 2\,m $ \\
 
 5. $ |E_i(\theta_e,\theta_{\gamma}) - E_i | < E_i $ \\
 
 6. $ \theta_i < 170^{\circ} $ \\

 7. $ 10^{\circ} < \theta_e < 160^{\circ} $  or
    $ 10^{\circ} < \theta_{\gamma} < 160^{\circ} $ \\
 
 8. $ 20 < (E-p_z)_{total} < 80\GeV$  \\
 
 9. $-30 < z$-vertex$ < 40\cm$, if $z$-vertex is reconstructed \\

10. $ m > 10\GeV $ \\
 
11. Visual scan \\ \hline
\end{tabular}
\caption{Selection requirements for the exclusive electromagnetic
cluster analysis.
Here $m$ is the invariant mass of the $e\,\gamma$ system.
The other quantities are explained in the text.
If more than two isolated e.m.\ clusters are found, the two
clusters fulfilling requirements 1 and 2 with the largest
polar angles $\theta_i$ are used.}
\label{tab.ls.sel.wab}
\end{center}
\end{table}

We define an electromagnetic cluster as one which has
at least $90\,\%$ of its energy
deposited in the electromagnetic part of the calorimeters and which is isolated.
The isolation requirement is 
based on a geometric cone in terms of azimuthal
angle ($\Delta \phi$) and pseudo-rapidity ($\Delta \eta$)
differences with respect to the center of the cluster. If less than
$10\,\%$ additional energy is found outside the cluster
but within 
$R = \sqrt{\Delta \phi^2 + \Delta \eta^2} < 0.5$, the cluster
is defined as isolated.

Requirement 2 takes into account that for most elastic
and quasi-elastic WAB events at low
$e\,\gamma$ masses
the summed energy of the two clusters is
given by the energy, $E$, of the initial electron.
Requirement 3 rejects events with other particles in the
final state and thus represents the
`empty detector' condition. Here $E_{total}$ is 
the total energy deposited in the BEMC and LAr
calorimeters.
Requirements 4 and 5 are loose cuts in order to ensure 
the proper kinematics: for exclusive $e^*$
($\rightarrow e\,\gamma$) production as well as for the
elastic and quasi-elastic WAB background the kinematics of
the process is over-constrained.
Using the polar angles $\theta_e$ and $\theta_\gamma$ alone,
the electron or photon energies
$E_i(\theta_e,\theta_{\gamma})$
or the mass $m(\theta_e,\theta_{\gamma})$
of the $e\,\gamma$ system can be calculated. 
These quantities
can then be compared to the measured energies, $E_i$, and to the
fully reconstructed invariant mass, $m$, using the measured
4-vectors.
Requirement 6 excludes a region where WAB background is especially large.
Requirements 7 and 8 suppress background from 
low squared momentum transfer 
Neutral Current (NC) Deep Inelastic
Scattering (DIS) and
$\gamma\,p$ events, respectively.
Requirement 9 suppresses events which are not due to positron proton
collisions.

442 events have passed the selection criteria including a
cut $m>10\GeV$ . These events were scanned visually.
Cosmic ray induced events (29 events) and beam halo muon events
(39 events) of complicated topology
were identified and removed at this
stage. 124 events did not satisfy the $e\gamma$ hypothesis,
because they had tracks pointing to two clusters or  
tracks uncorrelated with the electromagnetic clusters.
This latter class of events is  not due to the WAB process
as was confirmed by a scan of WAB Monte Carlo events~\cite{bib.epcompt}
corresponding to an integrated luminosity of $6\pb^{-1}$ .
The same scanning procedure was
also performed for selected Two Photon Monte Carlo events~\cite{bib.lpair}
(integrated luminosity of $9\pb^{-1}$).
The generated WAB and $\gamma\gamma$ events are processed
using the H1 detector simulation program and are then  
subject to the same reconstruction and analysis procedures  as the real
data.

The remaining 250 events are compared
to the WAB event simulation which predicts 
$257.2\,\pm\,13.7$ events
(for an integrated luminosity  
of $2.75\pb^{-1}$)
 and to the $\gamma\,\gamma$
event simulation, where $9.4\,\pm\,1.6$ events are
expected. 

\begin{figure}[htb]
\begin{center}
\epsfig{figure=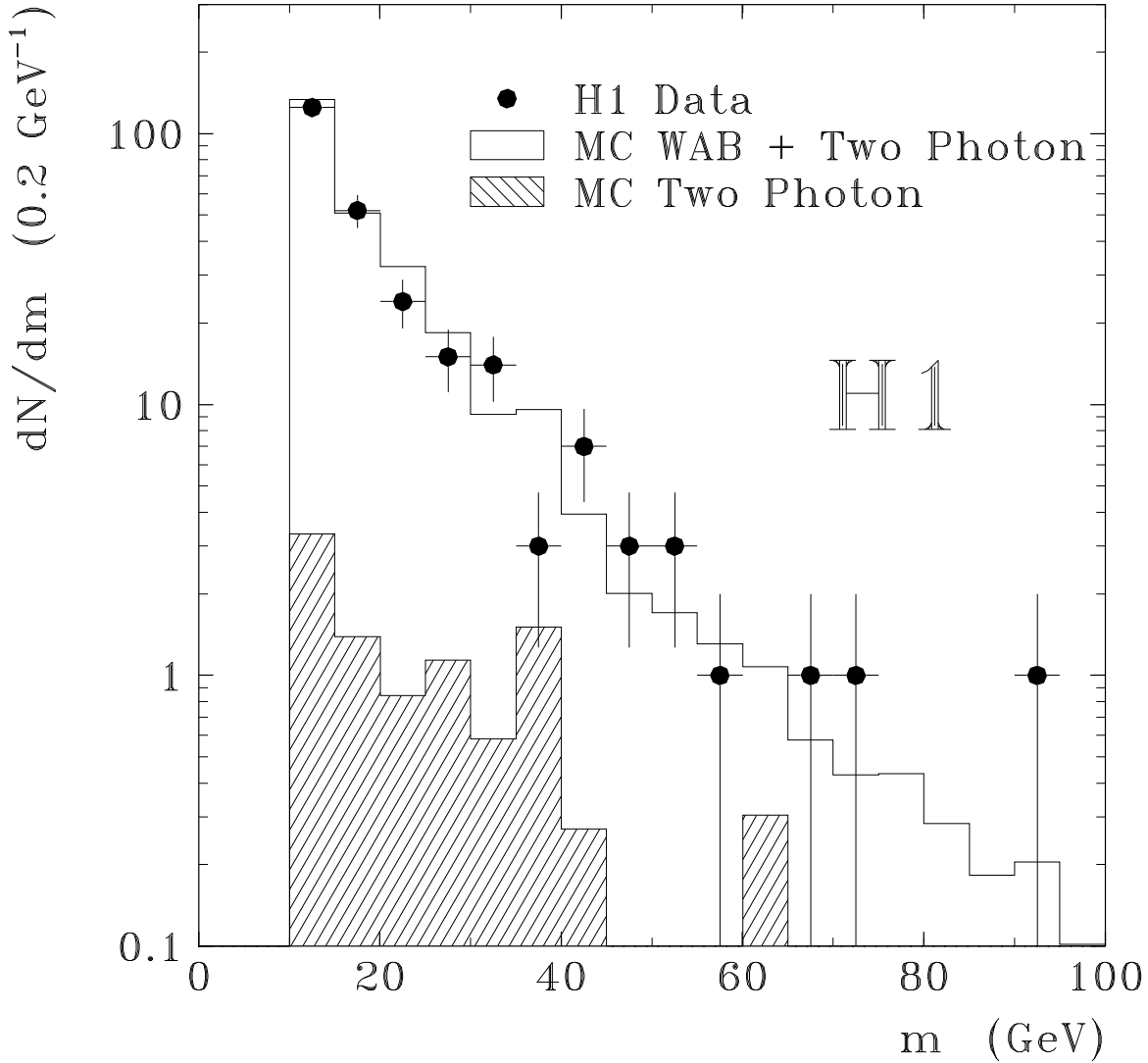,width=\textwidth}
\caption{Mass spectrum of the selected $e\,\gamma$ pairs with masses 
above $10\GeV$ in the exclusive electromagnetic cluster analysis. 
Only statistical errors are shown.
The shaded histogram shows the absolute prediction based on the 
Two Photon Monte Carlo simulation, and the open histogram 
shows the expectation for the sum of 
the Two Photon and WAB processes.}
\label{fig.ls.wab.mass}
\end{center}
\end{figure}

The  distributions
of the energies and polar angles of the electromagnetic clusters (not shown)
as well as the $e\,\gamma$ mass spectrum (figure~\ref{fig.ls.wab.mass})
are reproduced well both in shape and absolute value
by the WAB and $\gamma\gamma$ Monte Carlo simulations.
Applying the Kolmogorov-Smirnov test~\cite{bib.kolmo} to the mass spectrum 
the probability
is $90\,\%$ that both follow the same distribution.
The highest measured mass is at $91.3\GeV$. 
Above 50 GeV 4.6 events are expected and 4 events are seen.
 In total $0.6$ events are
expected from the WAB process for masses above $100\GeV$.

\subsection{Inclusive electromagnetic cluster analysis}

Candidates for more general event topologies from heavy lepton decays,
with two or more electromagnetic clusters as well as other particles in the
final state, must fulfill the selection criteria shown in
table~\ref{tab.ls.sel.highet}.
The selections in table~\ref{tab.ls.sel.highet}
are motivated by Monte Carlo studies
and will ensure high efficiencies for the high mass region in the channels
$e^* \rightarrow e\, \gamma$
and
$e^* \rightarrow e\, Z$, $Z \rightarrow e\,e$ for elastic and
inelastic $e^*$ production.

\begin{table}[htb]
\begin{center}

\begin{tabular}{l}
 \hline
 1. $ \geq 2$ isolated e.m.\ clusters, $E_{t1} > 20\GeV$,
       $E_{t2} > 10\GeV$ \\
 
 2. $ 20 < (E-p_z)_{total} < 80\GeV$  \\
 
 3. $-30 < z$-vertex$ < 40\cm$, if $z$-vertex is reconstructed \\

 4. Visual scan \\ \hline
\end{tabular}
\caption{Selection requirements for the inclusive electromagnetic
cluster analysis.
$E_{t1}$ and $E_{t2}$ are the transverse energies of the
first and second cluster.
If more than two isolated electromagnetic clusters are found, the two
clusters with largest transverse energies are used.}
\label{tab.ls.sel.highet}
\end{center}
\end{table}

328 events were selected and scanned visually and 65 cosmic muons 
and 167 beam halo muon events were removed from the data sample.
First, the decay channel 
$e^* \rightarrow e\, \gamma$
was analyzed.
75  events are 
due to the DIS process where
a hadronic jet near a $\phi$-crack of the LAr calorimeter was 
misidentified as an electromagnetic cluster. 
Two events are not considered as candidates
for the $e\,\gamma$ final state as they have tracks pointing to both
clusters. There remain 19 candidates for the decay channel
$e^* \rightarrow e\, \gamma$
which are compared to the WAB~\cite{bib.epcompt} and 
NC DIS~\cite{bib.lepto} event simulation. The expectations
from these two processes are $10.8\,\pm\,2.0$ and
$4.5\,\pm\,1.3$ events, respectively.

\begin{figure}[htb]
\begin{center}
\epsfig{figure=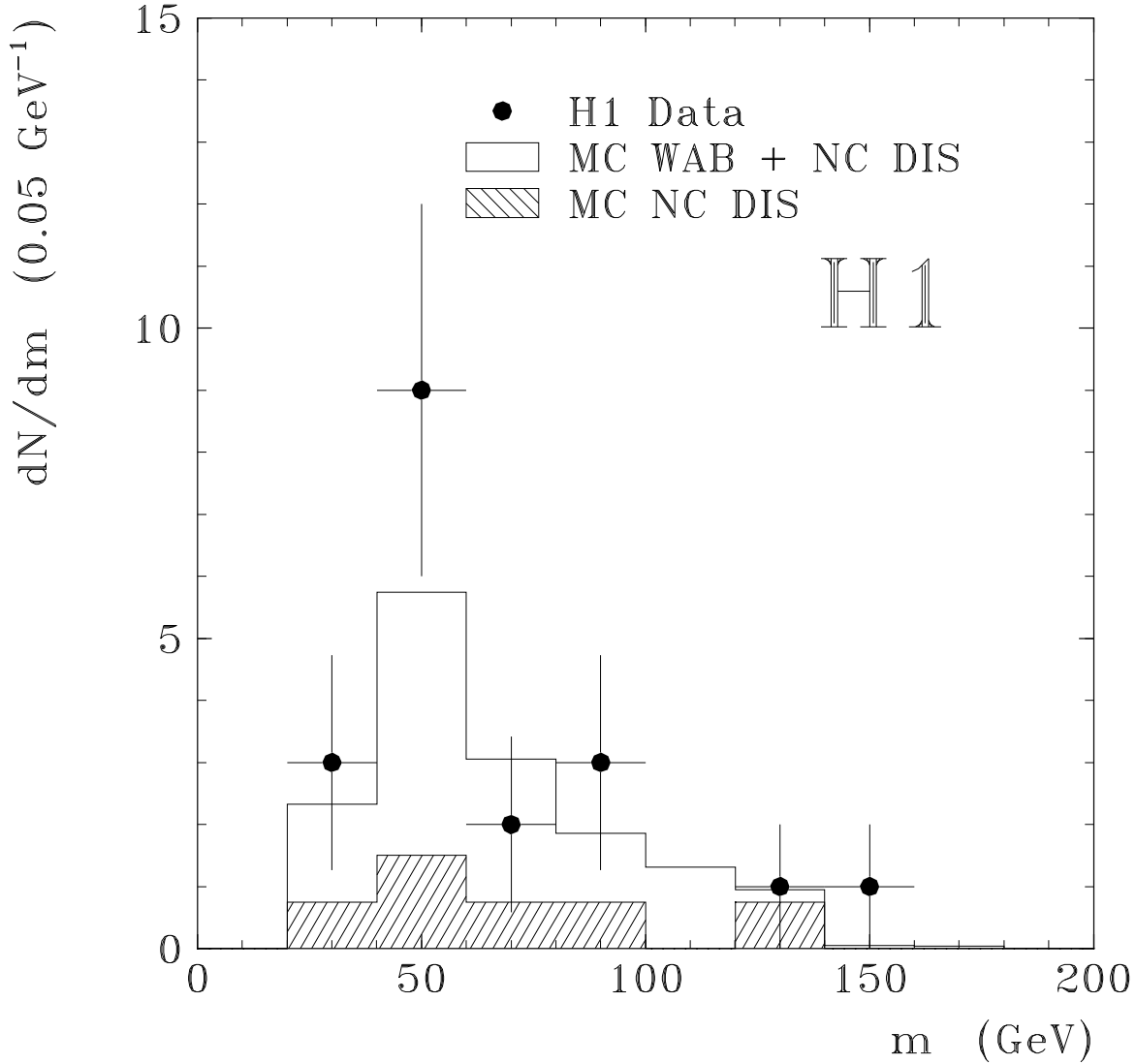,width=\textwidth}
\caption{Mass spectrum of the selected
high $E_{t}$ electromagnetic cluster pairs in the inclusive electromagnetic cluster analysis.
The shaded histogram shows the absolute prediction based on the 
DIS Monte Carlo simulation, and the open histogram 
shows the expectation for the sum of 
DIS and WAB reactions.}
\label{fig.ls.highet.mass}
\end{center}
\end{figure}

The absolute values and the shapes of the distributions
for the transverse energies and polar angles 
of the electromagnetic clusters (not shown)
as well as the invariant mass spectrum of the two clusters 
(figure~\ref{fig.ls.highet.mass})
are reproduced well by the Monte Carlo simulation.
Applying the Kolmogorov-Smirnov test to the mass spectra, the probability
is $69\,\%$ that both follow the same distribution.
The highest measured mass is at $154.3\GeV$. In total $0.06$ events are
expected from the WAB process for masses above $150\GeV$.

Candidates for the decay channel
$e^* \rightarrow e\, Z$, $Z \rightarrow e\,e$
(21 events) must fulfill  the following additional cuts:
 if there are exactly two
isolated electromagnetic clusters in the event, both must have a polar angle,
$\theta$, above $20^{\circ}$. The invariant mass of the two cluster
system must be between $80\GeV$ and $95\GeV$. If more than two clusters
are found, the third cluster must have $E_{t3} > 5\GeV$. 
The invariant mass, reconstructed from only two of the three clusters,
which is nearest to the mass of the Z boson is then required
to be between $80\GeV$ and $95\GeV$.

These additional selections are applied to all 96 previously selected
physics events (i.e. not cosmic or beam halo muon induced);
no event survives and
there are therefore no candidates left for this decay channel.

\subsection{Cluster and missing transverse energy analysis}
 
Events with at least one isolated high transverse energy 
electromagnetic cluster 
in addition to a large missing transverse energy are
candidates for the decay channels
$e^* \rightarrow e\,Z$, $Z \rightarrow \nu\,\nu$,
$e^* \rightarrow \nu\,W$, $W \rightarrow e\,\nu$,
or
$\nu^* \rightarrow \nu\,\gamma$ in both elastic and inelastic production
processes.
Background is expected from the NC DIS process with large fluctuations
of missing transverse energy.
Candidates for these event topologies must fulfill the selections 
shown in table~\ref{tab.ls.sel.clptm}.

Requirement 3 minimizes background from NC DIS reactions and enhances a
possible signal with highly  energetic neutrinos in the final state.
Requirement 5, which excludes events with a muon track, 
serves as a simple cosmic
and beam halo muon event filter.

\begin{table}[htb]
\begin{center}

\begin{tabular}{l}
 \hline
 1. $ \geq 1$ isolated electromagnetic cluster, $E_{t} > 20\GeV$ \\
 
 2. $E_{t}^{miss} > 20\GeV$ \\

 3. $ (E-p_z)_{total} < 50\GeV$  \\
 
 4. $-30 < z$-vertex$ < 40\cm$,  $z$-vertex must be reconstructed \\

 5. No rec.\ track in the instrumented iron with $ \geq 8$ hits \\

 6. Visual scan \\ \hline
\end{tabular}
\caption{Selection requirements for the electromagnetic
cluster and missing transverse energy analysis.
$E_{t}$ is the transverse energy of the cluster,
and $E_{t}^{miss}$ is the total missing transverse energy
of the event.
If more than one isolated electromagnetic cluster is found, that 
with largest transverse energy is used.}
\label{tab.ls.sel.clptm}
\end{center}
\end{table}

268 events were selected and scanned visually and 111 cosmic events
and 153 beam halo muon events were identified and removed at this
stage. In one event
a hadronic jet was misidentified as an electromagnetic cluster. Finally  three
candidate events remain, where $0.8\,\pm\,0.5$ events are expected
from NC DIS~\cite{bib.lepto} interactions.

For candidate events the 4-vector of the missing neutrino system,
and therefore the invariant mass of the excited lepton, can be reconstructed
for the decay channels
$e^* \rightarrow e\,Z$, $Z \rightarrow \nu\,\nu$
and
$\nu^* \rightarrow \nu\,\gamma$.
For the channel
$e^* \rightarrow \nu\,W$, $W \rightarrow e\,\nu$
the invariant mass of the two neutrinos was assumed to be that of the
$W$-boson. This choice is a compromise and overestimates
(underestimates) the reconstructed excited lepton mass in the low (high)
mass region.

The reconstructed excited lepton masses of the three surviving events are
$139\GeV$, $148\GeV$ and $186\GeV$ for the hypothesis
$e^* \rightarrow e\,Z$, $Z \rightarrow \nu\,\nu$, and
$131\GeV$, $134\GeV$ and $165\GeV$ assuming
$e^* \rightarrow \nu\,W$, $W \rightarrow e\,\nu$.
For both hypotheses no significant deviation in the mass distribution
is observed compared to
the NC DIS Monte Carlo simulation.

Events for the decay channel 
$\nu^* \rightarrow \nu\,\gamma$
are required  to satisfy the additional requirement that there is no
track with transverse momentum, $p_t$, above
$5\GeV$ inside a cone of $R=0.1$ in the $\eta$-$\phi$-plane,
centered around the isolated
cluster,  where $R$ is defined as in the previous analysis
(see section 3.1).
None of the three data events survives this additional selection.

\subsection{Inclusive muon analysis}
 
Events with high transverse momentum ($p_t$) muons
in the final state are selected
as candidates for the decay channels
$e^* \rightarrow e\,Z$, $Z \rightarrow \mu\,\mu$
and
$e^* \rightarrow \nu\,W$, $W \rightarrow \mu\,\nu$ of elastically and
inelastically produced excited leptons.
High energy muon candidates in the final state are to be expected from the
Two Photon process, $\gamma\,\gamma \rightarrow \mu\,\mu$,
from photo- and electro-production of $J/\Psi$-mesons,
from meson decays and
from hadronic punch-through.
The basic selection cuts are shown in table~\ref{tab.ls.sel.muon}.

\begin{table}[htb]
\begin{center}

\begin{tabular}{l}
 \hline
 1. $ \geq 1$ track in the instrumented iron \\
 
 2. Link to a central vertex \\ 

 3. $p_t > 10\GeV$ of a central track linked to the iron track\\
 
 4. Rejection of throughgoing cosmic ray muons \\

 5. Visual scan \\

 6.1 Second track in the instrumented iron, link to a central track \\

 6.2 Invariant mass of two muon system greater than $60\GeV$ \\

 7.1 No second track in the instrumented iron with link 
to a central track \\

 7.2 $\leq 2$ tracks beside muon track \\

 7.3 No central track opposite in polar and azimuthal angle
to the muon  track \\ \hline
\end{tabular}
\caption{Selection requirements for the inclusive muon analysis.
$p_t$ is the transverse momentum of the $z$-vertex fitted central track.
Candidates for the decay channels
$e^* \rightarrow e\,Z$, $Z \rightarrow \mu\,\mu$
or
$e^* \rightarrow \nu\,W$, $W \rightarrow \mu\,\nu$
have to satisfy selection criteria 1--6 or 1--5 and 7, respectively.}
\label{tab.ls.sel.muon}
\end{center}
\end{table}

After applying the cuts 1--4 shown in table~\ref{tab.ls.sel.muon},
170 events remained and were 
visually scanned. Of these, 147 events were 
found to be cosmic or beam halo muons,
leaving 16 single muon events, 6 two muon events, and one event with
four muons in the final state.
Applying the additional cuts 6.1 and 6.2 to select candidates for the 
decay channel
$e^* \rightarrow e\,Z$, $Z \rightarrow \mu\,\mu$,
no event remained. All masses of the two muon system are below $44\GeV$
and the events are therefore rejected.
As well as the standard selection criteria 1--5 candidates for the decay channel 
$e^* \rightarrow \nu\,W$, $W \rightarrow \mu\,\nu$
have to satisfy 
the additional cuts 7.1--7.3.  
Only elastic $e^*$ production is considered (cut 7.2)
because it is difficult to identify single muons within
jets.
No event survives the cuts in this channel.

\subsection{Jet analysis}

In the context of this paper events with a final state including 
two or more jets with a high invariant mass 
are  
candidates for decays of the heavy gauge bosons $W$ and $Z$.
Background events are expected from photoproduction, NC DIS, or from 
Charged Current (CC) DIS
processes.
The cuts to select candidates for the decay channels
$e^* \rightarrow e\,Z$, $Z \rightarrow q\,\bar{q}$
and
$e^* \rightarrow \nu\,W$, $W \rightarrow q\,\bar{q}'$
are shown in table~\ref{tab.ls.sel.jet}.

\begin{table}[htb]
\begin{center}

\begin{tabular}{l}
 \hline
 1. $ \geq 2$ jets, $E_{ti} > 15\GeV$, with $i=1$, 2 \\
 
 2. $m_{12} > 60\GeV$ \\

 3. $E_{tag} < 2\GeV$ \\

 4. $-30 < z$-vertex$ < 40\cm$, $z$-vertex must be reconstructed \\

 5. Visual scan \\

 6. Electron candidate, $\theta_e < 140^{\circ}$ \\

 7.1 Reject events with an electron candidate \\

 7.2 $E_t^{miss} > 20\GeV$ \\

 7.3 $(E-p_z)_{total} < 50\GeV$ \\ \hline
\end{tabular}
\caption{Selection requirements for the jet analysis.
$m_{12}$ is the invariant mass of the two jet system.
If more than two jets are found, the two jets with 
highest transverse energies $E_{ti}$ are used.
$E_{tag}$ is the energy deposited in the electron tagger.
An electron candidate is defined as an isolated  electromagnetic cluster
with $E_e > 10\GeV$ and $\theta_e > 10^{\circ}$.
The electron candidate with highest energy is ignored in  the
jet search.}
\label{tab.ls.sel.jet}
\end{center}
\end{table}

A cone algorithm with radius $R=1$ in
the $\eta$-$\phi$-plane is used to define jets.
The two jets with largest transverse energies are used.
If an electron candidate, defined as the most energetic isolated
electromagnetic cluster satisfying $E_e > 10\GeV$ and $\theta_e > 10^{\circ}$,
is found, it is  not included in the jet search.
Requirement 3 rejects photoproduction events in which the scattered
positron is detected in the electron tagger.
Requirement 4 suppresses beam-gas events.
The events are visually scanned and background events (cosmic muons)
as well as events with misidentified electron candidates or jets
are rejected. The same scanning procedure was  also performed for 
Monte Carlo events.

The events were required to have a  LAr transverse
energy trigger.
The efficiency of the LAr transverse energy trigger has been
estimated using 55 events satisfying cuts 1, 2, 4, 5, and 7.1, which 
were triggered by the electron tagger. Of these events 51 also have
a LAr transverse energy trigger, giving an  efficiency
of $93\,\%$. If both jets have transverse energies, $E_{ti}$, greater
than $20\GeV$, the trigger efficiency rises to $100\,\%$.

Candidates for the decay channel
$e^* \rightarrow e\,Z$, $Z \rightarrow q\,\bar{q}$
are required to fulfill selection criteria 1--6.
In this channel 15 data events are selected to be compared with
the expected 
$14.6\,\pm\,2.3$ events
from a NC DIS Monte Carlo simulation~\cite{bib.lepto}.

\begin{figure}[htb]
\begin{center}
\epsfig{figure= 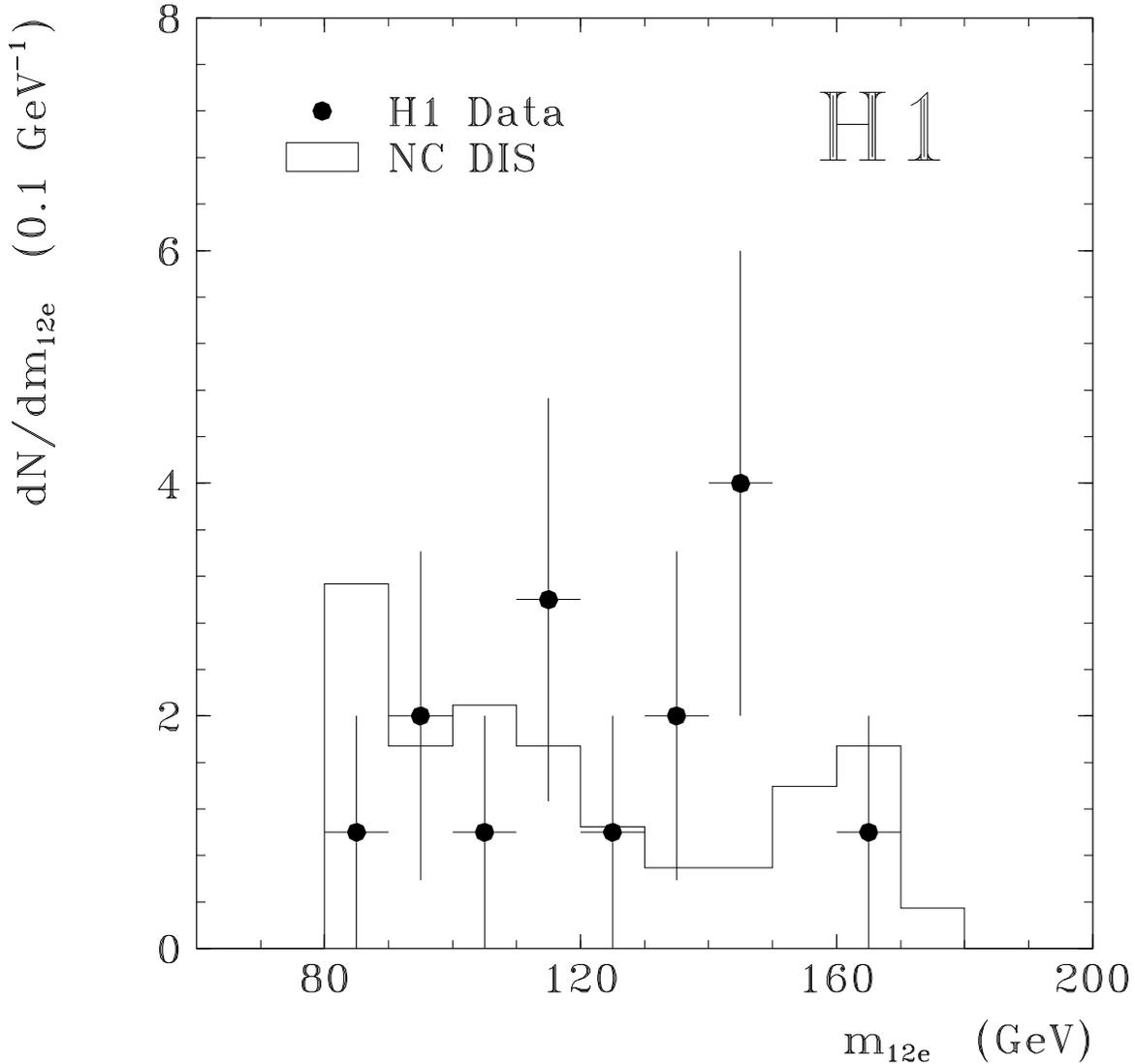,width=\textwidth}
\caption{Mass spectrum of the selected electron and two jets.
The histogram shows the absolute prediction based on the 
NC DIS Monte Carlo simulation.}
\label{fig.ls.jet.mass}
\end{center}
\end{figure}

The absolute number of events
as well as the invariant mass spectrum 
of the two jets and the electron, $m_{12e}$,
(figure~\ref{fig.ls.jet.mass})
are well reproduced by the Monte Carlo simulation.
Applying the Kolmogorov-Smirnov test to the mass spectra
(figure~\ref{fig.ls.jet.mass})
the probability is $43\,\%$ that both follow the same distribution.
The highest measured mass is at $m_{12e} = 164\GeV$.

Candidates for the second decay channel
$e^* \rightarrow \nu\,W$, $W \rightarrow q\,\bar{q}'$
are required to fulfill selection criteria 1--5 and 7.
Three data events survive the cuts compared with the $0.6\,\pm\,0.6$ events
expected from a Monte Carlo simulation of photoproduction~\cite{bib.pythia}.
A rough estimate shows that from the CC DIS process
approximately one further event is expected.

The invariant masses of the two jet system, $m_{12}$, for the three
events are $73\GeV$, $112\GeV$, and $75\GeV$ respectively.
The 4-vector of the missing neutrino is reconstructed and
the invariant masses of the two jet system and the neutrino,
$m_{12\nu}$ are
found to be $120\GeV$, $135\GeV$, and $141\GeV$ respectively.

In summary, after imposing selection requirements for 
heavy gauge bosons, no significant excess of events is found in any
of the channels under study.
Where candidates still survive the cuts, their number is compatible
with expectations from specific background processes.

\subsection{Results of $e^*$ and $\nu^*$ search}
 
Table~\ref{tab.ls.eff} shows the total  efficiencies, including that of the 
trigger,  for all excited lepton channels under study as a function of
excited lepton mass.
The Monte Carlo event generator COMPOS~\cite{bib.compos}
based on the cross section calculated in ref.~\cite{Hagiwara}
is used.
We employ the Lund string model~\cite{bib.lund} for hadronization
and decays. All results presented here have been obtained using
the MRS~H~\cite{bib.mrsh} parameterization of the parton densities.
The generated events are passed through the full H1 detector simulation
and are reconstructed in the same manner as the real data.

\begin{table}[htb]
\begin{center}
\begin{tabular}{ll|rrrrrrrrr}
 \hline
&&\multicolumn{9}{c}{Mass $\ell^*$/GeV}\\
\multicolumn{2}{c|}{Decay channel} & 25 & 50 & 75 & 81 & 92 & 100 &
  150 & 200 & 250 \\
 \hline
$e^* \rightarrow e\,\gamma$ & &
  46 & 73 & 83 &--- &--- & 91 & 90 & 88 & 89 \\
$e^* \rightarrow e\,Z$, & $Z \rightarrow e\,e$ &
     &    &    &    & 67 & 63 & 84 & 82 & 75 \\
$e^* \rightarrow e\,Z$, & $Z \rightarrow \mu\,\mu$ &
     &    &    &    & 53 & 55 & 54 & 31 & 23 \\
$e^* \rightarrow e\,Z$, & $Z \rightarrow \nu\,\nu$ &
     &    &    &    &  0 &  0 & 84 & 87 & 80 \\
$e^* \rightarrow e\,Z$, & $Z \rightarrow q\,\bar{q}$ &
     &    &    &    &  2 & 27 & 77 & 69 & 64 \\
$e^* \rightarrow \nu\,W$, & $W \rightarrow e\,\nu$ &
     &    &    & 80 &--- & 82 & 79 & 82 & 79 \\
$e^* \rightarrow \nu\,W$, & $W \rightarrow \mu\,\nu$ &
     &    &    & 28 &--- & 30 & 11 & 4 & 6 \\
$e^* \rightarrow \nu\,W$, & $W \rightarrow q\,\bar{q}'$ &
     &    &    &  0 &--- &  8 & 64 & 55 & 44 \\
$\nu^* \rightarrow \nu\,\gamma$ & &
  13 & 24 & 47 &--- &--- & 58 & 60 & 61 & 71 \\ \hline
\end{tabular}
\caption{Total analysis efficiencies for different decay
channels of the excited leptons. Efficiencies are
given as percentages and are based on 200 simulated Monte Carlo
events for each entry of the table, from which an absolute efficiency error
of 2--3\,\% follows. An entry `---' means that no events
were generated at this mass.}
\label{tab.ls.eff}
\end{center}
\end{table}

\begin{figure}[htb]
\begin{center}
\epsfig{figure=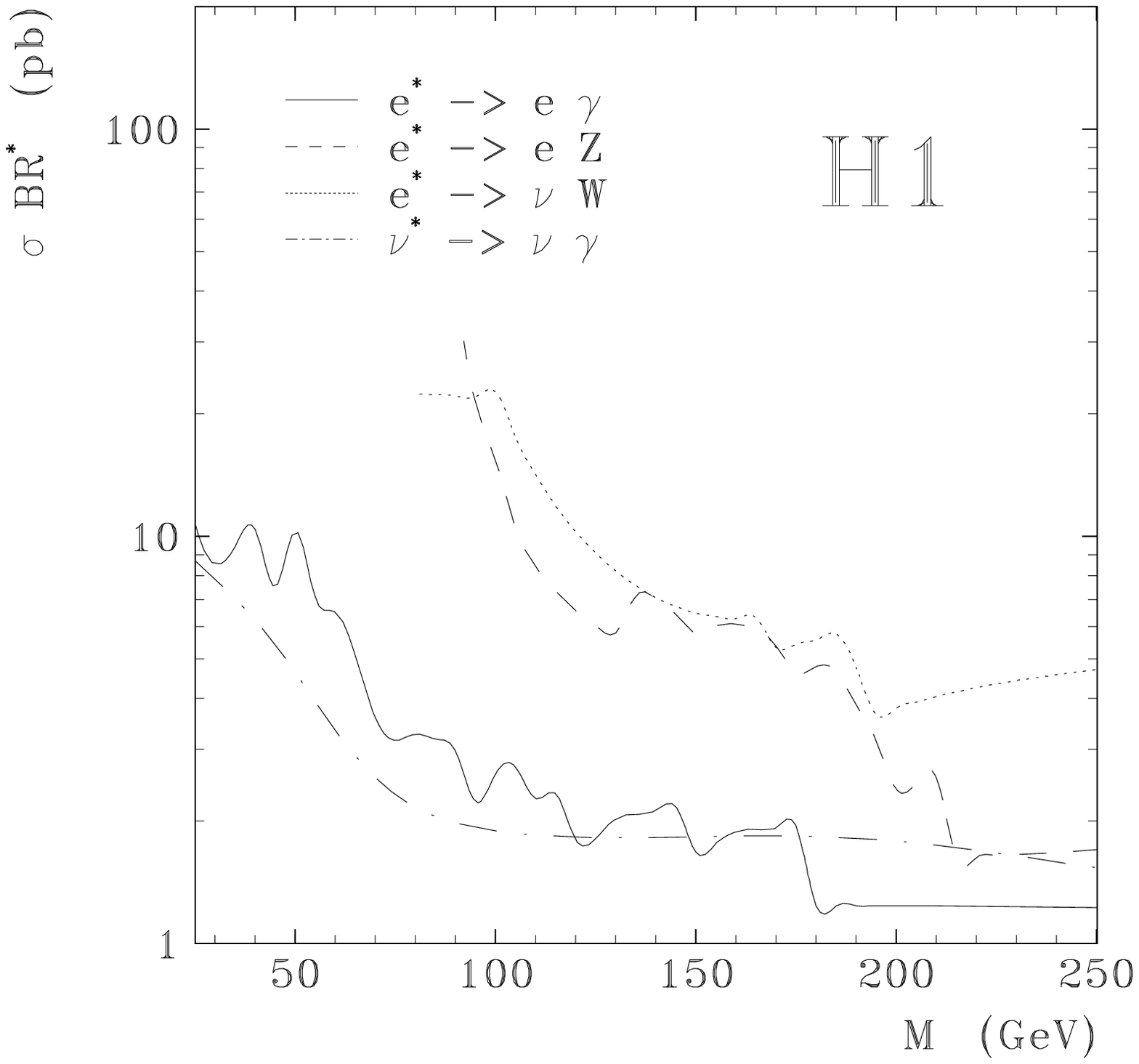,width=\textwidth}
\caption{Rejection limits with a CL of $95\,\%$ for 
$e^*$ and $\nu^*$. Regions above the curves are excluded.
Decay channels of the heavy gauge bosons $W$ and $Z$ are
combined.}
\label{fig.ls.lim94}
\end{center}
\end{figure}

The experiment can set limits on the product $\sigma\cdot BR^*$ 
of the production
cross section with the branching ratio into
a specific decay channel.
The limits are derived by following the procedure recommended
by the Particle Data Group~\cite{bib.holger}.
Figure~\ref{fig.ls.lim94} shows the rejection limits at a $95\,\%$
Confidence
Level (CL) for the $e^*$ and $\nu^*$ as a function of 
the masses of the excited states.
In calculating limits for the channel $e^* \rightarrow e\,\gamma$
both samples from the exclusive and the inclusive electromagnetic cluster
analysis are combined as are the 
different decay channels of the $W$ and $Z$ heavy gauge bosons. 
The rejection limits for final states with a $W$ or $Z$ boson are 
similar, the main difference stemming from $Z$ decays into
$\nu\,\bar{\nu}$.
The results improve earlier published H1 results~\cite{
bib.h1_94}
by a factor 3 to 6 in the limits for the excluded cross section.

In the following the uncertainties of the limits 
on $\sigma \cdot BR^*$
for the channel
$e^* \rightarrow e\,\gamma$
are discussed. The wiggles are due to possible statistical 
fluctuations of the number of candidate events.
The other decay channels have fewer candidate events and therefore little
or no  fluctuations of the calculated limits. The uncertainties due to
efficiency and integrated luminosity  are similar for all channels.

A shift of the calorimeter energy scale  by
$\pm 5\,\%$ effectively shifts both the event masses 
and the resulting limits by approximately $\pm 5\,\%$
on the horizontal axis.
A variation of sensitive energy cuts in the analysis
by $\pm 5\,\%$ leads to statistical limit fluctuations
of 5--10\,\% for $e^*$ masses below $100\GeV$,
and has no effect for higher $e^*$ masses.

The statistical uncertainty on the background estimator
of $10\,\%$ in the lowest mass bin, the error in the knowledge of the total
efficiency (2--3\,\%), and the error in the integrated luminosity (1.5\,\%) 
lead to a 15\,\% limit variation 
for $e^*$ masses below $100\GeV$,
which decreases to 3.3\,\% at high $e^*$ masses, where
no candidates were found. In addition the use of other 
parton density parameterizations leads to variations in the derived
limits of up to $10\%$.

Using the specific model for excited leptons
of Hagiwara et al.~\cite{Hagiwara} one can also calculate
limits on the quantity 
$c/{\sl \Lambda} \cdot \sqrt{BR^*}$, 
where the coupling constant $c \equiv c_{\gamma e^* e}$ for $e^*$ and
$c \equiv c_{W \nu^* e}$ for $\nu^*$.
For the channel $e^* \rightarrow e\,\gamma$
and $e^*$ masses of $50\GeV$, $100\GeV$, \ldots, $250\GeV$
the corresponding limits (in TeV$^{-1}$) with $95\,\%$ CL are
1.21, 0.97, 1.34, 2.45, and 11.1.
For a typical value $c_{\gamma e^* e}^2 = 1/4$ and an $e^*$ mass
of $100\GeV$ we currently rule out compositeness scale parameters
${\sl \Lambda} < 440\GeV$.

The $\nu^*$ production cross section 
is very much suppressed in positron proton collisions compared
to electron proton collisions
due to the different charge of the exchanged $W$ boson
and the different quarks probed inside the proton.
A derivation of an exclusion limit for $\Lambda$ is only
meaningful if the total width of the excited fermion is small compared
to its mass.  In the model of Baur, Spira and Zerwas~\cite{Zerwas2}
the decay of an excited neutrino into $\nu\gamma$ is allowed
if the couplings $f,f'$ are set to $1,-1$, which results in
$c^2_{\gamma\nu^*\nu}=1/4$. With this choice of parameters
$\Lambda < 19.5\GeV$ is excluded at a mass 
of $135\GeV$ where the total width already reaches 28 GeV.
For  
a $\nu^*$ mass of $100\GeV$ compositeness
scale parameters ${\sl \Lambda} < 51\GeV$
are ruled out by this experiment.


Searches for excited leptons are also performed in
$e^+e^-$ annihilation experiments~\cite{bib.l3_95}.
At present the results obtained are restricted to masses
below $\sim 135$ GeV, however with better limits on the coupling
than achieved in this experiment. The L3 collaboration e.g. quotes 
limits ($95\,\%$ CL)
on $c_{\gamma e^* e}/{\sl \Lambda} \cdot \sqrt{BR^*}$
which range between 0.32--0.51$\TeV^{-1}$ for $e^*$ masses from
90--130\GeV.

\section{Search for excited quarks}

\subsection{Cluster and jet analysis}

As mentioned in the introduction the $q^*$ production is dominated by 
$\gamma$ exchange with very small values of the squared momentum transfer
and the scattered positron is nearly always unseen in the detector. So
the final state of $q^* \rightarrow q\,\gamma$ events is characterized by 
an isolated electromagnetic cluster with large transverse energy,
$E_t$, and a jet which balances this $E_t$.
Therefore the main backgrounds which could mimic
the signal are
NC DIS events for which the electron track is lost (mismeasured),
WAB events
with the electron identified as a jet,
and photoproduction events with a high 
$E_t$ jet having a large enough electromagnetic component to simulate 
a photon, or with the production of a prompt photon.
The cuts to select candidates for the channel $q^* \rightarrow q\,\gamma$
are shown in table~\ref{tab.qs.sel}.

\begin{table}[htb]
\begin{center}

\begin{tabular}{l}
 \hline
 1. $ \geq 1$ isolated e.m.\ cluster, $E_{t\,\gamma} > 20\GeV$ \\

 2. $ 20^{\circ} < \theta_{\gamma} < 160^{\circ} $ \\

 3. No track pointing to the e.m cluster \\

 4. No electron candidate \\

 5. $ \geq 1$ jets, $\Delta p_{x,y} < 3\sigma $ \\

 7. \ $z$-vertex must be reconstructed \\


 8. Visual scan \\ \hline
\end{tabular}
\caption{Selection requirements for the cluster and jet analysis.
The variables are explained in the text.
If more than one isolated electromagnetic cluster is found, that
with the largest transverse energy is used as the photon candidate.
The jet giving the best balance to the photon  
in transverse energy
is used.}
\label{tab.qs.sel}
\end{center}
\end{table}

Requirements 1 to 3 are the basic selection cuts for photon candidates.
The photon is taken as the highest $E_t$ electromagnetic shower found
in the LAr calorimeter with no corresponding track, i.e.  no
 vertex-fitted track within a cone of
$R=\sqrt{ (\eta_{tr} - \eta_{em} )^2 + (\phi_{tr} - \phi_{em} )^2 } < 0.3$
around the electromagnetic cluster center 
and no non-vertex-fitted central track with a projection 
on the front face of the calorimeter closer than $10\cm$
to the electromagnetic cluster center.  
These cuts reject most of the NC DIS events.

An independent search for electrons is then performed (requirement 4).
Events with both a photon and an electron are rejected.
This criterion mainly rejects  WAB events
and NC DIS events with  wide angle final state  radiation.

For the definition of the jets a cone algorithm with radius
$R = 1$ in the $\eta$-$\phi$-plane is used.
At least one jet is requested (requirement 5). In the case of more
than one jet being found, the jet which gives the best $E_t$ balance
to the electromagnetic cluster is kept.
A $3\sigma$ $p_x$ and $p_y$ balance cut is then applied. This last
selection is  mass dependent and for each event the value of $\sigma$ is chosen 
according to the invariant mass of the photon-jet-system.
The dependence of 
the $\Delta p_x$ and $\Delta p_y$ distributions widths on the $q^*$ mass
has been determined using Monte Carlo  events generated with such a
$q^*$ decay.


After applying all selection cuts, 8 candidates for 
$q^* \rightarrow q\,\gamma$
are found. A visual inspection of these candidates allows a clear
interpretation of 6 of them : 5 are NC DIS with the electron track 
badly reconstructed and one is a photoproduction event.

\begin{figure}[htb]
\begin{center}
\epsfig{file=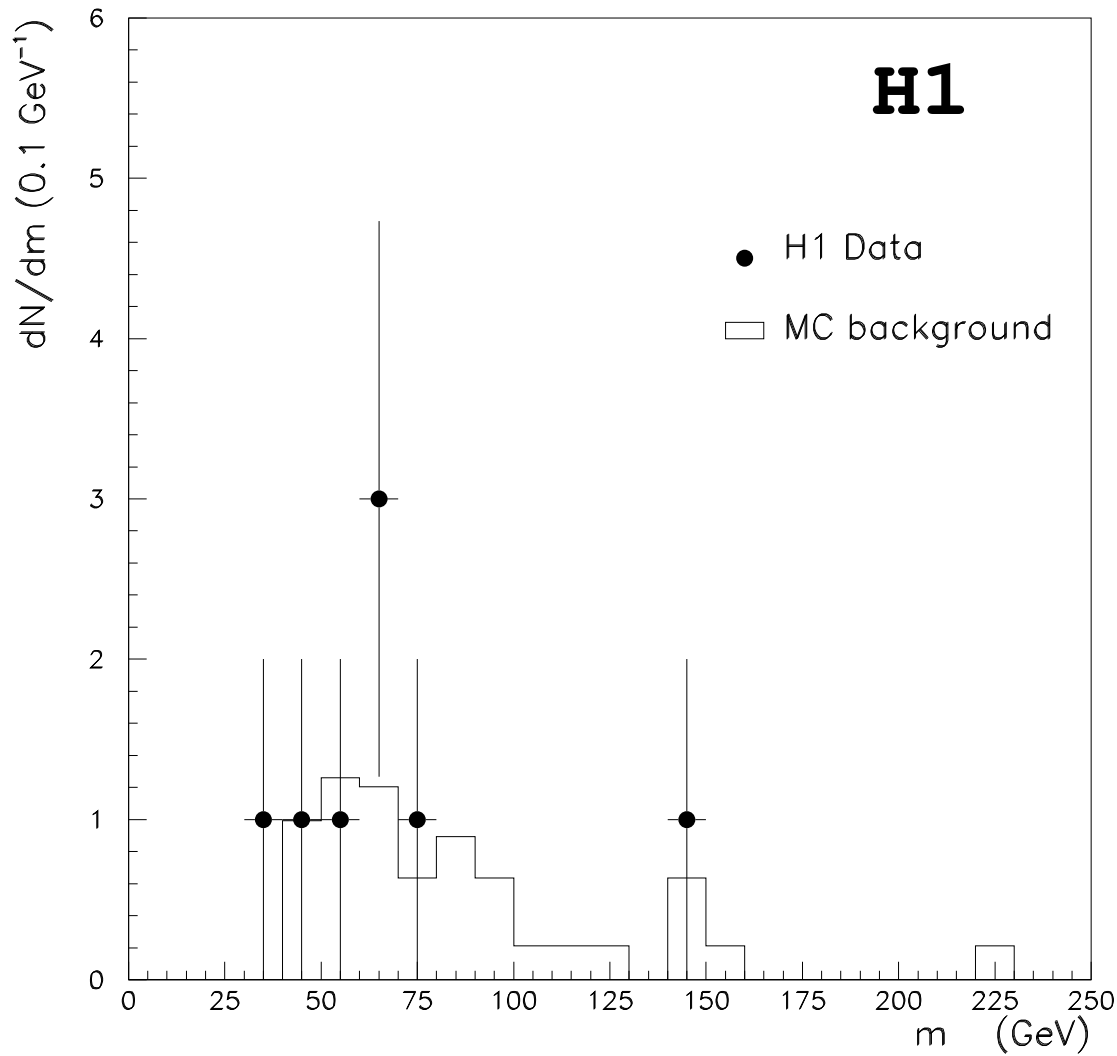,width=\textwidth}
\caption{Invariant mass spectrum of the photon-jet-system.
The points correspond to the data and the histogram to the 
background expectation.}
\label{fig.qs.mass}
\end{center}
\end{figure}

For the background estimation, the generated NC DIS Monte Carlo events, corresponding 
to an integrated
luminosity of $26.9\pb^{-1}$, the WAB events~\cite{bib.courau} ($18.4\pb^{-1}$)
and the photoproduction events (resolved: $3.72\pb^{-1}$, direct:
$4.04\pb^{-1}$, direct with charm: $4.12\pb^{-1}$)
are all passed through the H1 detector simulation program and are 
subject to the same reconstruction and analysis procedure as the real
data.
The expected number of events from these different background processes,
normalized to the integrated luminosity of $2.75\pb^{-1}$,
are $4.03 \pm 1.12$, $1.25 \pm 0.65$ and $2.04 \pm 1.18$ respectively.
The quoted errors include statistical errors and an error of $1.5\,\%$
on the integrated luminosity .
The invariant mass of the photon-jet-system of 
the 8 events is shown on Figure~\ref{fig.qs.mass}.
The experimental mass resolution varies from 3.5 to 10 GeV for a
$q^*$ mass between 50 to 250 GeV. 
The absolute number of the events and the shape of the distribution  
are well described by the Monte Carlo simulation.

\subsection{Results of $q^*$ search}

No model assumption is made 
in the search for a $q^*$ decaying into a quark and
a photon. 
However,  in order to give limits on cross
sections, specific assumptions are used to determine the
detection efficiency for $q^*\rightarrow q\gamma$ events. The
Monte Carlo event generator COMPOS~\cite{bib.compos} is  used
with the Lund string model giving  fragmentation and
decays and  
the MRS~H~\cite{bib.mrsh} parameterization of parton densities.
The generated events are then passed through the full H1 detector
simulation and reconstructed in the same manner as the real data. 
The total detection 
efficiency is shown in table~\ref{tab.qs.eff}, as a function
of the $q^*$ mass.

\begin{table}[htb]
\begin{center}
\begin{tabular}{c|rrrrrr}
 \hline
&\multicolumn{6}{c}{Mass $q^*$/GeV}\\
Decay channel & 50 & 75 & 100 &
  150 & 200 & 250 \\
 \hline
$q^* \rightarrow q\,\gamma$ &
  30 & 47 & 50 & 55 & 49 & 53 \\ \hline
\end{tabular}
\caption{Total detection efficiency, given as a percentage based on 1000
simulated Monte Carlo events for each mass.}
\label{tab.qs.eff}
\end{center}
\end{table}

\begin{figure}[htb]
\begin{center}
\epsfig{file= 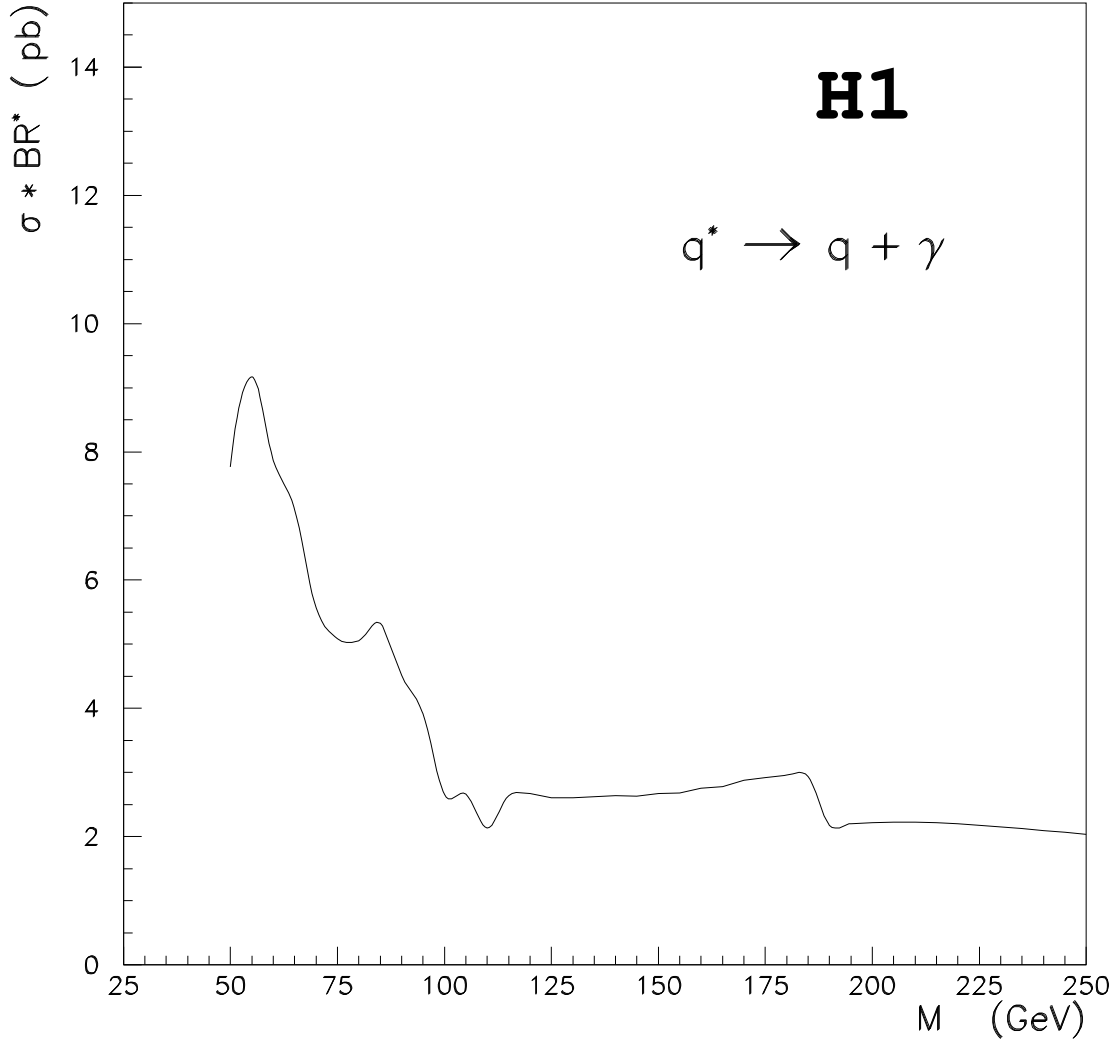,width=\textwidth}
\caption{Rejection limit with a CL of $95\,\%$ on
$\sigma (q^*) \cdot BR(q^* \rightarrow q + \gamma)$.
The region above the curve is excluded.}
\label{fig.qs.lim}
\end{center}
\end{figure}

Finally
the $95\,\%$ CL rejection limit on 
$\sigma (q^*) \cdot BR(q^* \rightarrow q + \gamma)$ 
is shown in figure~\ref{fig.qs.lim} 
as a function of the $q^*$ mass.
Cross sections above a value of about $9\pb$ for a $q^*$ mass of $50\GeV$ down to
a value of $2\pb$ for a $q^*$ mass of $250\GeV$ are ruled out at the $95\,\%$ CL.

A variation of $\pm 2\,\%$ on the electromagnetic and $\pm 5\,\%$ 
on the hadronic energy scale of the LAr calorimeter corresponds to
a systematic error on the limit which decreases from $+35\,\%$ for 
a $q^*$ mass of $50\GeV$
(due mainly to a larger rejection rate with the $p_t$ selection), 
to  $10\,\%$ for a $q^*$ mass above $75\GeV$.

To determine the influence of the  specific model used in computing detection 
efficiencies,
we have simulated an isotropic angular decay distribution by 
re-weighting the COMPOS events. The efficiency for finding the signal 
decreases from $19\,\%$ to $33\,\%$ for a $q^*$ mass 
in the range 50--250 $\GeV$. This would correspond to the limit for  
$\sigma (q^*) \cdot BR(q^* \rightarrow q + \gamma)$ varying from 11 to
$3\pb$.

Assuming a specific model~\cite{Zerwas2} our limit can be 
transformed into a limit on the 
compositeness scale, $ \Lambda $. However,  this limit is strongly
dependent on the relative  coupling of $q^* \rightarrow q V$
(where $V$ stands for $ \gamma \  Z  \  W $) to $ q^* \rightarrow q g $.
As an example, for a $ q^* $ mass of 100 GeV 
the limit on $\Lambda$ goes  
from about
60 GeV ( if the BR($  q^{*} \rightarrow q g $) $ \simeq 95 \% $ ) to 290 GeV
( if the BR($ q^{*} \rightarrow q g $) $ \simeq 25 \% $ ).
Our results are complementary to a search in proton anti-proton
collisions~\cite{bib.95_CDF} which investigates $q^*$ production
via quark gluon excitation.

\section{Conclusions}
Using data taken in 1994 new limits for the production  of
excited electrons in high energy electron proton interactions
have been  obtained. For excited electrons the search includes
the decay channels $e^*\rightarrow e\gamma$, $e^*\rightarrow eZ$ and
$e^*\rightarrow \nu W$, with the heavy bosons decaying into leptons
or jets. New limits extending up to
masses of 250 GeV are obtained. 
A search for excited neutrinos  is also
included considering only electromagnetic decays. In
addition new limits for $q^*$ production via electromagnetic
excitation and decay are presented which complement
results obtained at hadron colliders.

{\bf Acknowledgments}
We are grateful to the HERA machine group whose
outstanding efforts made this experiment possible. We appreciate the immense
effort of the engineers and technicians who constructed and maintain
the H1 detector. We thank the funding agencies for financial support.
We acknowledge the support of the DESY technical staff. We wish to
thank the DESY directorate for the support and hospitality extended to
the non-DESY members of the collaboration.

\end{document}